\documentstyle[prl,aps,twocolumn,psfig,epsf,floats]{revtex}
\begin{document} 
 
\title{A Superconducting Persistent Current  Qubit} 
 
\author{ 
T. P. Orlando$^{\mbox{\scriptsize a}}$, 
J. E. Mooij$^{\mbox{\scriptsize a,b}}$, 
Lin Tian$^{\mbox{\scriptsize c}}$, 
Caspar H. van der Wal$^{\mbox{\scriptsize b}}$, 
L. Levitov$^{\mbox{\scriptsize c}}$, 
Seth Lloyd$^{\mbox{\scriptsize d}}$, 
J. J. Mazo$^{\mbox{\scriptsize a,e}}$ 
} 
 
\author{\footnotesize 
$^{\mbox{\scriptsize a}}$Dept.\ of  
Electrical Engineering \& Computer Science, 
M.I.T., Cambridge, MA, 02139, USA\\ 
$^{\mbox{\scriptsize b}}$Dept.\ of Applied Physics, Delft University of 
Technology, P.O.Box 5046, 2628 CJ Delft, The Netherlands\\ 
$^{\mbox{\scriptsize c}}$Dept.\ of Physics, 
M.I.T., Cambridge, MA, 02139, USA\\ 
$^{\mbox{\scriptsize d}}$Dept.\ of Mechanical Engineering, 
M.I.T., Cambridge, MA, 02143, USA\\ 
$^{\mbox{\scriptsize e}}$ Departamento de F\'{\i}sica  
de la Materia Condensada and ICMA, CSIC-Universidad de 
Zaragoza, E-50009 Zaragoza, Spain\\ 
} 
\date{\today} 
\maketitle

\abstract{We present the design of a superconducting 
qubit  that has circulating currents of opposite 
sign as its two states. The circuit consists of 
three nano-scale aluminum Josephson junctions connected in 
a superconducting loop and controlled by magnetic fields. 
The advantages of this qubit are that it can be  
made insensitive to background charges in the substrate, 
the flux in the two states can be detected with a SQUID, and 
the states can be manipulated with magnetic fields.  
Coupled systems of qubits are also discussed as well 
as sources of decoherence. 
} 

\section{Introduction}

Quantum computers are devices that store information on quantum 
variables such as spins, photons, and atoms, and that process 
that information by making those variables interact in a way 
that preserves quantum coherence  
\cite{Lloyd93,Bennett95,DiVincenzo95,SpillerIEEE,Lloyd_SciAm95}. 
Typically, these variables consist of two-state  
quantum systems called quantum bits or  
`qubits' \cite{Schumacher95}.   
To perform a quantum computation, 
one must be able to prepare qubits in a desired initial state, 
coherently manipulate superpositions of a qubit's two states, 
couple qubits together, measure their state, and keep 
them relatively free from interactions that induce noise 
and decoherence  
\cite{Lloyd93,Bennett95,DiVincenzo95,SpillerIEEE,Lloyd94,Landauer}. 
Qubits have been physically implemented in a variety of 
systems, including cavity quantum  
electrodynamics\cite{Turchette},  
ion traps\cite{Monroe},  
and nuclear spins\cite{Gershenfeld,Cory}. 
Essentially any two-state quantum system that can be 
addressed, controlled, measured, coupled to its neighbors and  
decoupled from the environment, is potentially useful  
for quantum computation and quantum  
communications\cite{Lloyd95b,Deutsch95}.   
Electrical systems which can be produced by modern 
lithography, such as nano-scaled quantum dots and tunnel 
junctions, are attractive candidates for constructing qubits:  
a wide variety of potential designs for qubits and their 
couplings are available, and the qubits are easily scaled 
to large arrays which can be integrated in electronic circuits 
\cite{DiVincenzo95,Kane98}.   
For this reason mesoscopic superconducting circuits 
of ultra-small Josephson junctions have been proposed as 
qubits\cite{Mooij-talk,Science_paper,Bocko97,Shnirman97,Averin98} 
and we detail one such circuit in  
this paper.     
 
Compared with the photonic, atomic, and nuclear qubits already  
constructed, solid state proposals based on lithography such as  
the one described here have two considerable disadvantages and  
one considerable advantage.  The first disadvantage is noise  
and decoherence 
\cite{DiVincenzo95,Lloyd94,Landauer}: 
the solid state environment has a higher density of states 
and is typically more strongly coupled to the degrees of  
freedom that make up the qubit than is the environment for 
photons in cavities, ions in ion traps, and nuclear spins 
in a molecule or crystal.  Extra care must be taken in solid 
state to decouple the qubit from all sources of noise and 
decoherence in its environment.  The second disadvantage 
is manufacturing variability\cite{Landauer}: each ion in an 
ion trap is identical by nature, while each lithographed 
Josephson junction in an integrated circuit will have  
slightly different properties.  Solid state designs must 
either be insensitive to variations induced by the manufacturing 
process, or must include a calibration step in which the 
parameters of different subcircuits are measured and compensated 
for \cite{Kane98}. 
 
The advantage of solid state lithographed circuits is their 
flexibility: the layout of the circuit of Josephson junctions 
or quantum dots is determined by the designer, and its parameters 
can be adjusted continuously over a wide range.  As the results 
presented in this paper demonstrate, this flexibility allows 
the design of circuits in which the variables that register 
the qubits are only weakly coupled to their 
environment.  In addition, the flexibility in circuit 
layout allows many possible options for coupling 
qubits together, and for calibrating and adjusting the qubits'  
parameters.  That is, the advantage of flexibility in design 
can compensate for the disadvantages of decoherence and  
manufacturing variability.     
 
The flexibility in design afforded by lithography conveys a  
further advantage to constructing quantum computers.  As noted 
above, a qubit has  to accomplish at least five functions:  
it has to be 
addressed, controlled, measured, coupled to its neighbors and 
decoupled from the environment.    
One of the axioms of design, 
%whether it is design of quantum computers, conventional computers, 
%or automobiles,  
is that the number of parameters that characterize 
a system's design should be at least as great as the number of 
parameters that characterize the system's function\cite{Suh}.   
%Simplicity is not necessarily a virtue:  if a 
%design is too parsimonious, then the design parameters typically  
%cannot be adjusted to insure that the system actually accomplishes 
%its different functions.  
The problem of having too few design 
parameters available is particularly acute in  
the design of quantum  
computers and qubits : a quantum computer is  
a device in which essentially  
every physical degree of freedom is used to register information  
and to perform the computation.  Degrees of freedom that are not 
used to compute are sources of noise and must be isolated from the 
computing degrees of freedom.  Designs for quantum computers 
are accordingly more constrained by fundamental physics than are 
designs for conventional computers: if one is storing information 
on a cesium atom, then the ``design parameters'' of  
the cesium atom ---its energy  
levels, decoherence times, interaction strengths, etc.  
---are fixed by nature once and for all.   
In the lithographed Josephson junction 
circuits proposed here, by contrast,  
it is possible to make qubits that have 
a variety of different design parameters, each of which can be 
adjusted to optimize different functions.

\section{Josephson-Junction Qubits}

The superconducting Josephson tunnel  
junction is described by a critical current 
$I_o$ and a capacitance $C$. (We will assume 
that the resistive channel of the junction is 
negligibly small.) For superconducting circuits the geometrical 
loop inductance $L_s$ is also important 
if $\Lambda = L_J/L_s< 1$, where $L_J=\Phi_o/2\pi I_o$ is 
the inductance associated with a Josephson junction in the loop. 
Here $\Phi_o=h/2e$ is the superconducting flux quantum. 
Josephson circuits can be divided into two general 
categories. Circuits of the first type have 
$\Lambda \gg 1$ so that the induced flux 
in the loop is not important. These circuits are typically made 
of aluminum, and the mesoscopic nature of their electronic 
transport has been studied in nanoscaled circuits. 
Circuits of the second type have $\Lambda \ll 1$ and 
induced flux caused by circulating currents is important. 
These circuits are typically made of niobium, and  
the macroscopic nature of the tunneling of flux  
has been studied in small-scaled circuits.

The prospects of using  superconducting  
circuits of the first type  as  qubits is encouraging 
because extensive experimental and theoretical work 
has already  been done on mesoscopic superconducting circuits. 
(For a review of this work see Chapter~7 in \cite{Tinkhambook} 
and in Ref.~\cite{STCbook}.) 
In circuits of the first type ($\Lambda \gg 1$), 
two energy  
scales determine the quantum mechanical  behavior: 
The Josephson coupling energy, $E_J=I_o \Phi_o/2\pi$, 
and the coulomb energy for single charges, $E_c=e^2/2C$. 
The energies can be  determined by the phases  of the  
Cooper pair wave function of  the nodes (islands) and the number of 
excess Cooper pairs on each node.  
The phase and the number can be expressed as 
quantum mechanical conjugate   
variables\cite{Curruthers68}. 
 
In the ``superconducting'' limit $E_J>E_c$, the phase is 
well defined and the charge fluctuates. In the ``charging'' 
limit, the charges on the nodes are well defined and the 
phase fluctuates strongly. When $E_J$ and $E_c$ are 
within a few orders  of magnitude of each other, the eigen 
states must 
be considered as quantum mechanical superpositions of 
either charge states or phase states. Such superposition 
states are important in designing  qubits. 
Experimental studies have been performed by several groups  
with aluminum tunnel junctions with dimensions below 
100\thinspace nm\cite{Tinkhambook,STCbook}.  
Superposition of charge states in circuits in the charging regime 
have been demonstrated\cite{Joyez94,Matters95,Bouchiat97} 
and are in  
quantitative agreement with theory\cite{Averin91,Lafarge96}. 
The Heisenberg uncertainty principle has been demonstrated when 
$E_J\approx E_c$\cite{Elion94,Matters95}. 
When $E_J>E_c$ topological excitations known as vortices 
exists and quantum mechanical interference of these 
quantities has been observed\cite{Ouden96}. 
Unfortunately  circuits of the first type in the 
charging regime are sensitive  to fluctuating 
off-set charges that are present in the substrate 
\cite{SET_noise,SET_noise2}.  
These 
random off-set charges make difficult the design 
of a controllable array of quantum circuits and introduce 
a strong source of decoherence. 
 
In circuits of the second type ($\Lambda \ll 1$), the quantum variables 
can be related to the flux in the loops and their  
time derivatives. For a superconducting loop with a   
single Josephson junction, known as an  
rf SQUID,  
thermal activation of macroscopic quantum states\cite{Han93} 
has been observed as well as  
macroscopic quantum tunneling between  states  
and  the discrete  nature of the quantum states\cite{Rouse95}. 
One of the advantages of these rf SQUID  systems is that 
the two  states have circulating currents of  
opposite sign and, hence, produce a readily measurable 
flux of opposite signs. 
To date no superposition of states have been demonstrated in 
the niobium circuits, although the  improving quality of 
the niobium  tunnel junctions may allow such 
a demonstration\cite{Diggins95,Castellano96}.

The goal of this paper is to design a qubit  
using circuits of the first type with aluminum, yet 
to have states (like in circuits of the second type) 
that are  circulating currents 
of opposite sign. These circulating current states 
create a magnetic flux of  
about $10^{-3}\Phi_o$ and we refer to these as 
``persistent current (PC) states.''  
These states 
obey all five functional requirements for a quantum bit: 
(1)~The superconducting circuit is at sufficiently a low  
temperature that the the PC states can be made insensitive to background charges  
and effectively decoupled from their electrostatic environment. 
The magnetic coupling to the  PC states and the environment can also be
made sufficiently weak.

\section{The Circuit} 
 
\begin{figure}[t] 
%\hspace{3.5in} 
\epsfysize=2.25in 
\epsfbox{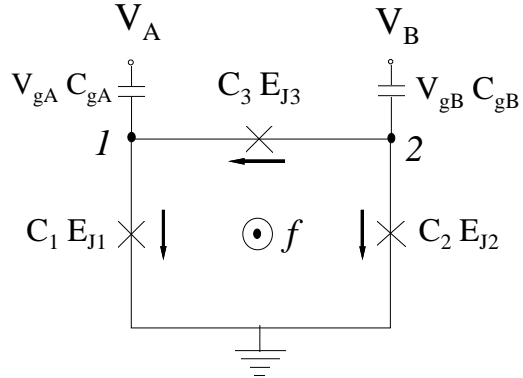} 
     \caption{The three-junction  qubit.   
Josephson junctions~1 and 2  both have 
Josephson energies $E_J$ and capacitance $C$ and  
Josephson junction~3 has a  
Josephson energy and capacitance $\alpha$ times 
larger. 
The nodes 1 and 2 represent 
the superconducting islands (nodes) 
which are coupled by gate capacitors $C_g= 
\gamma C$ to gate voltages $V_A$ and $V_B$. 
The arrows define the direction of the currents.  
The flux is taken out of the page.} 
  \label{fig_qubit3} 
  \end{figure}

The circuit of the qubit is shown in Fig.~\ref{fig_qubit3}. 
Each junction is  marked by an ``x'' 
and is modeled\cite{Tinkhambook,ODbook} 
by a parallel combination of an ideal 
Josephson junction and a capacitor $C_i$. The parallel 
resistive channel is assumed negligible. 
The  ideal Josephson junction has a  
current-phase relation, $I_i=I_o \sin\varphi_i$ where $\varphi_i$ is 
the gauge-invariant phase of junction $i$.  
 
For the calculation of the energy 
the inductance of the loop is considered negligible  
$\Lambda \gg 1$ so that the 
total flux is the external flux. In this case, 
fluxoid  quantization around the loop containing the junctions, gives 
$\varphi_1 - \varphi_2 + \varphi_3= -2\pi f$. 
Here $f$ is the magnetic frustration and is the amount of external 
magnetic flux in the loop in units of the flux quantum $\Phi_o$. 
 
The Josephson energy due to each junction is $E_{Jn}(1-\cos\varphi_n)$. 
The total Josephson energy $U$ is then 
$  U = \sum_i E_{Ji} (1-\cos{\varphi_i} )$. Combined with the flux 
quantization condition the Josephson energy  
is\cite{Doniach84} 
   \begin{equation} 
      \frac{U}{E_J}=  
2  +\alpha  - \cos\varphi_1 -\cos\varphi_2 -\alpha \cos(2\pi f +\varphi_1 -  
\varphi_2)\,. 
  \label{threejun_U} 
   \end{equation} 
 
\begin{figure}[t] 
%\hspace{2.25in} 
\epsfysize=2.25in 
\epsfbox{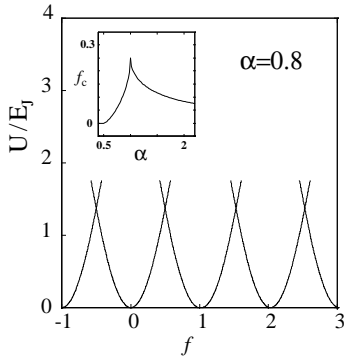} 
         \caption{$U/E_J$ vs. $f$ for $\alpha=0.8$ and for  
minimum energy  phase  configuration. 
The energy is periodic with period $f=1$ and is symmetric about 
$f=1/2$. Near $f=1/2$, there is a region $[1/2-f_c,1/2+f_c]$ where 
there are two stable solutions. The inset plots $f_c$ as a function of 
$\alpha$.  
              } 
       \label{fig_Eclass} 
      \end{figure}

The important feature of this Josephson   
energy is that it is a function 
of two phases \cite{Lik-L}.   
For a range of magnetic frustration $f$, 
these two phases, $\varphi_1$ and $ \varphi_2$, 
permit two stable  configurations which  
correspond to dc currents flowing in opposite directions. 
We illustrate this in Fig.~\ref{fig_Eclass}, 
where we plot the energy of the minimum of the system 
as a function of $f$ for $\alpha=0.8$. 
The energy is periodic with period $f=1$ and is symmetric about 
$f=1/2$. Near $f=1/2$, there is a region $[1/2-f_c,1/2+f_c]$ where 
there are two stable solutions.  
The inset plots $f_c$ as a function of 
$\alpha$.  
These two solutions  have circulating currents  
of opposite direction and are degenerate at $f=1/2$.  
The calculation of the  energy for the stable solutions 
and $f_c$ is given in Appendix~A.

The main feature of the  
qubit that is proposed in this paper is to use these 
two states of opposite current as the two states of the 
qubit. 
By adding the charging energy (the capacitive energy) of 
the junctions and considering the circuit quantum mechanically, 
we can adjust the parameters of the circuit  
so that the two lowest states of the system near $f=1/2$ 
will correspond to these two classical states of opposite circulating 
currents.  Moreover, we will show that these two states can 
be made  insensitive to the gate voltages and the random off-set  
charges.  The  quantum 
mechanics of the circuit will be considered in detail  in the next section.

\begin{figure}[t] 
%\vspace{3.5in} 
\epsfxsize=3.5in 
\epsfbox{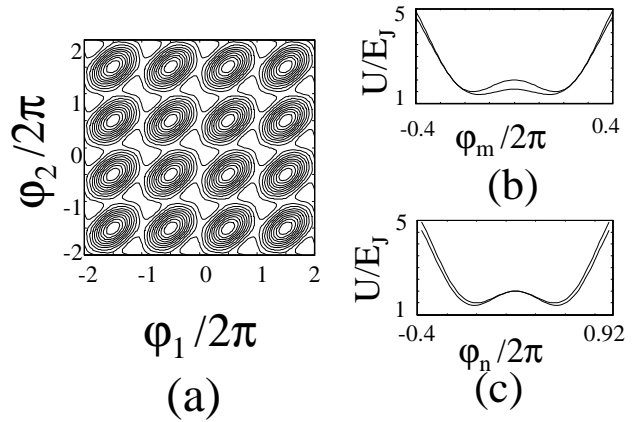} 
         \caption{(a)~A contour plot of the Josephson energy  
(potential energy) $U(\varphi_1,\varphi_2)$ for  $f=1/2$ for 
$\alpha=0.8$. 
The nested nearly circular shapes mark the maxima in the potential,
and the figure-eight shaped  contours enclose two minima.
(b)~A plot of 
the potential vs.~$\varphi_m$, the phase  along the direction  
between these two minimum in the same unit cell, 
(c)~A plot of the potential  
vs.~$\varphi_n$,  
the phase along  direction from one minima to its next nearest
neighbor.
Note that the barrier is a saddle point. 
The upper  curve in each figure is for $\alpha=1.0$ 
and the lower for $\alpha=0.8$. 
          } 
\label{fig_pot} 
      \end{figure}

The stable classical solutions correspond to energy minima in 
$U(\varphi_1,\varphi_2)$. Let's consider the case of $f=1/2$. For 
$\alpha \le 1/2$, $U$ has only one  minimum at 
$\varphi_1=\varphi_2=0 \bmod{2\pi}$.  Above the critical value of 
$\alpha=1/2$, this minimum bifurcates into two degenerate 
minima at  
$\varphi_1=-\varphi_2= \pm \varphi^*  \bmod{2\pi}$ where 
$\cos{\varphi^*}= 1/2\alpha$. The  minima  form a two-dimensional 
pattern with the two minima  at  
$(\varphi^*,-\varphi^*)$ and $(-\varphi^*,\varphi^*)$  
repeated in a  two-dimensional square lattice. 
This pattern can be seen in 
Fig.~\ref{fig_pot} which is a contour plot of  
the Josephson energy as a function of the phase variables for $\alpha=0.8$. 
The nested nearly circular contours   mark the maxima in the potential.
The figure-eight shaped  contour encloses two minima.
Fig.~\ref{fig_pot}b shows  
the potential along  $\varphi_m$, the phase  between the two minimum in 
a unit cell; that is, along the line $\varphi_2=-\varphi_1$.
The upper curve is for $\alpha=1.0$   and  the lower for 
$\alpha=0.8$.
Fig.~\ref{fig_pot}c shows the potential vs 
$\varphi_n$,
which connects one minimum (say at $(-\varphi^*,\varphi^*)$)
to its next nearest neighbor (at $(\varphi^*, 2\pi-\varphi^*)$).
For $\alpha=0.8$ the energy barrier between the  
two minima is much lower than the 
energy barrier from the minimum in one unit cell to the neighboring  
unit cell. For $\alpha=1.0$ the energy barrier from unit cell 
to unit cell is nearly the same as the barrier within the unit cell. 
The ability to manipulate the potential landscape by changing 
$\alpha$ will  be important 
in designing the qubit. 
 
We now consider  
the electric  energy $T$ stored in the five capacitors 
in the circuit.  Each capacitor  $C_j$ has  
a voltage  across it of $V_j$ so that 
  \begin{equation} 
T= \frac{1}{2}\sum_j C_j V_j^2 - Q_{gA}V_A -Q_{gB} V_B \,.     
   \end{equation} 
Here $j=1,2,3$ and ${gA}$ and $gB$. 
The last two terms subtract the work done by the voltage source 
to give the  
available electric (free)  energy\cite{ref-energy}. 
The voltage across each  Josephson junction is given by 
the Josephson voltage-phase relation 
$V_n=(\Phi_o/2\pi)\dot{\varphi}_n$, where  the over-dot indicates 
a partial time derivative. The ground in the circuit labels the 
zero of potential and is a virtual ground.

The voltage across the gate capacitor $gA$ is 
$V_{gA}=V_{A}-V_1$ and similarly for $V_{gB}=V_{B}-V_2$. 
The electric energy can then be written in terms of the 
time derivatives of the phases as 
   \begin{equation} 
T= \frac{1}{2}\left(\frac{\Phi_0}{2 \pi} \right)^2 
  \vec{\dot{\varphi}}^T\cdot {\bf C} \cdot \vec{\dot{\varphi}}\,. 
  \end{equation} 
The constant term 
$-\frac{1}{2}\vec{V}_g^T \cdot {\bf C}_g \cdot \vec{V}_g$ 
has been neglected and  
  \begin{equation} 
 \vec{\dot{\varphi}}=  \left( \matrix{ \dot{\varphi}_1 \cr 
                               \dot{\varphi}_2 \cr} 
  \right) 
\qquad 
  {\bf C} = C \left( 
                 \matrix{1 + \alpha + \gamma & -\alpha\cr 
                            -\alpha & 1 + \alpha + \gamma \cr} 
            \right) 
  \end{equation} 
and 
  \begin{equation} 
 \vec{V}_g= \left( \matrix{ V_{A} \cr 
                               V_{B} \cr} 
  \right) 
\qquad 
  {\bf C}_g = \gamma C \left( 
                 \matrix{ 1  & 0\cr 
                          0&   1  \cr} 
            \right) \,. 
   \end{equation} 
 
The classical equations of motion can be found from the 
Lagrangian ${\cal{L}}=T-U$. We take  the  
electrical energy as the kinetic energy 
and the Josephson energy as the potential energy 
\cite{L_circuit}.The canonical momenta is  
${P}_i=\partial{{\cal{L}}}/\partial \dot{\varphi_i}$. 
To attach a more physical meaning to the canonical momentum, 
we shift the Lagrangian by a Galilean-like transformation to 
   \begin{equation} 
{\cal{L}}= T - U - \left(\frac{\Phi_o}{2\pi} \right) 
\vec{\dot{\varphi}}^T\cdot  {\bf C}_g \cdot \vec{V}_g\,. 
    \end{equation} 
The canonical momentum is then 
    \begin{equation} 
   \vec{P}=  \left(\frac{\Phi_o}{2\pi} \right)^2 {\bf C} 
\cdot \vec{\dot{\varphi}} - 
 \left(\frac{\Phi_o}{2\pi} \right) {\bf C}_g \cdot \vec{V}_g 
    \end{equation} 
and is directly proportional to the charges at the  
islands at nodes $1$ and $2$ in Fig.~\ref{fig_qubit3}  
as 
   \begin{equation} 
  \vec{Q}= \frac{2\pi}{\Phi_o} \vec{P} \,.     
    \end{equation} 
(For any Josephson circuit it can be shown that  
there exist    linear combinations of the phases 
across the junctions such that these linear combination can be 
associated with each node and the corresponding 
conjugate variable is proportional 
to the charge at that  
node\cite{devoret-lecture,orlando-matrix}. 
If self and mutual inductances are need to be  included  
in the circuit (as we argue does not need to be done in 
our case), then 
additional conjugate pairs would needed\cite{orlando-matrix}.) 
 
The classical Hamiltonian, $H=\sum_i P_i\dot{\varphi}_i -{\cal{L}}$, 
is 
   \begin{equation} 
H= \frac{1}{2} \left(  
\vec{P} + \frac{\Phi_o}{2\pi}\vec{Q^I} \right)^T 
 \cdot {\bf M}^{-1}\cdot \left( \vec{P}  
+ \frac{\Phi_o}{2\pi}\vec{Q^I} \right)\nonumber \\ 
%& &+\frac{1}{2}\vec{V}_g^T \cdot {\bf C}_g \cdot \vec{V}_g 
 + U(\vec{\varphi}) 
    \end{equation} 
where the  effective mass  
${\bf M}=(\Phi_o/2\pi)^2 {\bf C}$ is anisotropic 
and the induced charge on 
the island is 
$\vec{Q}^I={\bf C}_g\cdot\vec{V}_g$. 
When driven by an additional  external current source, 
the classical dynamics of this system have been 
studied in recent years  both  
theoretically\cite{Yukon95,Geigenmuller96} 
and experimentally\cite{Barahona97,Caputo97}

Note that the kinetic energy part of this Hamiltonian is 
   \begin{equation} 
T= \frac{1}{2} \left( \vec{Q} + \vec{Q^I} \right)^T 
 \cdot {\bf C}^{-1}\cdot \left( \vec{Q} + \vec{Q^I} \right) 
%-\frac{1}{2}\vec{V}_g^T \cdot {\bf C}_g \cdot \vec{V}_g 
    \end{equation} 
which is just the electrostatic energy written is terms of 
the charges and induced charges on the islands. 
Often this is the method used in discussing the  
charging part of the Hamiltonian. See for example Reference 
\cite{devoret-lecture} and the references therein. 
A characteristic charge is $e$ and characteristic 
capacitance is $C$ so that the characteristic  
electric energy is the so-called charging energy, $E_c=e^2/2C$.

\section{Quantum Circuit} 
 
The transition to treating the circuit quantum mechanically is to 
consider the classically conjugate variables in the classical 
Hamiltonian as quantum mechanical operators\cite{Schoen88,Spiller92}.   
For example, the momenta 
can be written as $P_1=-i\hbar\partial/\partial \varphi_1$ and 
$P_2=-i\hbar\partial/\partial \varphi_2$ and the wave function can 
then be considered as $|\Psi>=\Psi(\varphi_1,\varphi_2)$. 
 
In this representation the plane-wave solutions, such as 
$\psi=\exp\{-i(\ell_1 \varphi_1 + \ell_2\varphi_2)\}$ 
corresponds to a state that has $\ell_1$ Cooper pairs on 
island (node) $1$  and $\ell_2$ Cooper pairs on 
island $2$. 
These plane-wave states are the so-called 
charging states of the  
system\cite{Kikharev85,Averin91}. 
Since a single  measurement of the  
number of Cooper pairs  on each island must 
be an integer, then so should the $\ell$'s here.  
(Note the expectation  value of the number of  
Cooper pairs is not restricted to an integer.) 
Furthermore, an eigen function $\Psi(\varphi_1, \varphi_2)$ 
can be written as a weighted linear combination of 
these charge states. 
This means that $\Psi(\varphi_1, \varphi_2)$ is periodic 
when each of the phases are changed by $2\pi$, as in the  
physical pendula\cite{pendularef}. 
 
By considering  
$\Psi(\varphi_1,\varphi_2)= 
\exp\{i(k'_1 \varphi_1 + k'_2\varphi_2)\} 
\chi(\varphi_1,\varphi_2) $ with 
$[k'_1,k'_2]=- (\gamma C/2e)[V_A,V_B]$, 
the Hamiltonian for  
$\chi(\varphi_1,\varphi_2)$ is almost the same but  
the induced charges are now transformed out of the problem, 
and we refer to this new Hamiltonian as the transformed Hamiltonian $H_t$, 
where\cite{H_footnote} 
   \begin{eqnarray} 
H_t= & &\frac{1}{2}  \vec{P}^T \cdot {\bf M}^{-1}\cdot \vec{P}  
+ E_J\{2  +\alpha  - \cos\varphi_1 -\cos\varphi_2  
 \nonumber \\ 
& & -\alpha \cos(2\pi f +\varphi_1 - \varphi_2) \} \,. 
    \end{eqnarray} 
The resulting equation   
$ H_t\chi(\varphi_1,\varphi_2)=E \chi(\varphi_1,\varphi_2) $ 
is the same as for an anisotropic, two-dimensional particle in the  
periodic potential $U$.  
The solutions are Bloch waves with 
the ``crystal momentum'' ${\bf k}$-values corresponding to 
${\bf -k'}$, which is proportional to the applied voltages. 
This choice of crystal momentum insures that  
$\Psi(\varphi_1,\varphi_2)$ is periodic in the phases.

We will first present the numerical results of the  
energy levels and wave functions for the circuit. 
Then we will use the tight-binding-like approximation 
to understand  the results semi-quantitatively. 
 
The eigen values and eigen wave functions for the transformed 
Hamiltonian $H_t$ are  solved numerically  by expanding the 
wave functions in terms of  states of constant charge  or states of 
constant phase. The states of constant charge result in the 
standard central equation for Bloch functions and are computationally 
efficient when $E_c > E_J$. The states of constant phase are solved 
by putting the phases on a discrete lattice and the numerics 
are more efficient when $E_J > E_c$. Since the Josephson 
energy dominates, we will show results computed  using  
the constant phase states. (However, when we used the 
constant charge states, we obtained the same results.)

The numerical  calculations are done in a rotated  
coordinate system which diagonalizes the capacitance matrix 
${\bf C}$ by choosing as coordinates the sum and difference of the phases, 
$\varphi_p= (\varphi_1 + \varphi_2)/2$ 
and $\varphi_m= (\varphi_1 - \varphi_2)/2$. The resulting 
reduced Hamiltonian is 
   \begin{eqnarray} 
H_t= & & 
 \frac{1}{2}\frac{ P_p^2}{M_p}+ \frac{1}{2}\frac{ P_m^2}{M_m} + 
E_J\left\{ 2  +\alpha \right.  \nonumber \\ 
& &\left. - 2\cos\varphi_p \cos\varphi_m 
               - \alpha\cos(2\pi f + 2\varphi_m) 
\right\} 
  \label{H_t_pm} 
   \end{eqnarray} 
where 
the momenta can be written 
as $P_p=-i\hbar\partial/\partial \varphi_p$  
and $P_m=-i\hbar\partial/\partial \varphi_m$. 
The mass terms are  
$M_p=(\Phi_o/2\pi)^2 2C(1+\gamma)$ 
and $M_m=(\Phi_o/2\pi)^2 2C(1+2\alpha +\gamma)$. 
In this coordinate system the full wave function 
$\Psi(\varphi_p,\varphi_m)= 
\exp\{i(k'_p \varphi_p + k'_m\varphi_m)\} 
\chi(\varphi_p,\varphi_m) $ with 
$[k'_p,k'_m]=- (\gamma C/2e) [V_A+V_B,V_A-V_B]$ 
and $ H_t\chi(\varphi_p,\varphi_m)=E \chi(\varphi_p,\varphi_m) $. 
Also the  
two minima of the potential $U(\varphi_p,\varphi_m)$ within a  
unit cell form a periodic  two-dimensional centered cubic lattice 
with lattice constants 
${\bf a}_1=2\pi {\bf i}_x$ and ${\bf a}_2=\pi {\bf i}_x + \pi {\bf i}_y$.

    \begin{figure}[t] 
\epsfxsize=3.3in 
\epsfbox{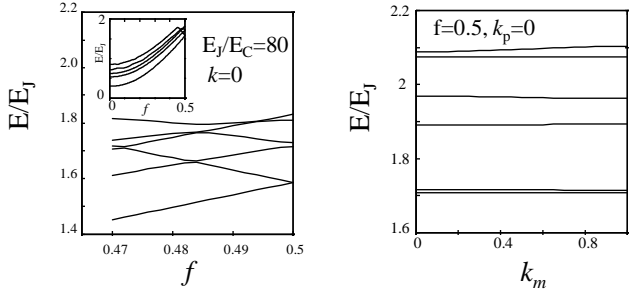}%{ek.eps} 
         \caption{The energy levels $E$ vs.~frustration and  
gate voltage for  
$E_J/E_c=80$,  $\alpha=0.8$, and $\gamma=0.02$. 
The gate voltage is related to the ${\bf k}$ values by 
$[k_p,k_m]=(\gamma C/2e) [V_A+V_B,V_A-V_B]$, 
(a)~$E/E_J$ vs $f_b$ near $f_b=1/2$ for $[k_p,k_m]=[0,0]$,  and 
(b)~$E/E_J$ vs $k_m$ for $k_p=0$. 
          } 
\label{fig_ebands} 
      \end{figure} 
 
Fig.~\ref{fig_ebands} shows the energy levels as a function 
of $f$  
and as a function  of the  gate voltage which is given in terms of 
${\bf k}$.  
We have taken $E_J/E_c=80$,  $\alpha=0.8$, and $\gamma=0.02$ 
in this  example. 
The  energy levels are symmetric about  $f=1/2$. 
In Fig.~\ref{fig_ebands}a, we see that  the two lowest  
energy levels near $f=1/2$, have opposite slopes, indicating 
that the circulating currents are of opposite sign.  
We also see that 
there is only  a small range of $0.485<f<0.5$ where the qubit 
can be operated between these states of opposite circulating current. 
This range is consistent with the range 
$[\frac{1}{2} \pm f_c]$
from the classical 
stability as shown in Fig.~\ref{fig_Eclass}. 
At $f=0.49$ 
direct calculation of the average circulating current, 
$<\Psi|I_o \sin\varphi_1|\Psi>$ gives 
that the circulating current for  the lower 
level is $I_1/I_o=-0.70$ and for the upper level is $I_2/I_o=+0.70$.  
(A  calculation of the circulating current from the 
thermodynamic relation 
$-\Phi_o^{-1}\partial E_n/\partial f$ gives the same result.) 
For a loop of diameter of $d=10\,\mu{\rm m}$,  
the loop inductance 
is of the order  
$\mu_o d \approx 10\, {\rm pH}$\cite{Zant94}. 
For $I_o\approx 
400 \, {\rm nA}$ ( $E_J = 200 \, {\rm GHz} $ ),  
the flux due to the circulating current
is $L I_1 \approx 10^{-3}\Phi_o$, 
which is detectable by an external SQUID. Nevertheless, the  
induced flux is small enough, that we are justified in  
neglecting its effect when calculating the energy levels.

The difference in energy between the lower and upper level at the 
operating point of $f=0.485$ is about $0.1\,E_J \approx 20\,{\rm GHz}$. 
Moreover, Fig.~\ref{fig_ebands}b shows that the energies of these 
levels is very insensitive  
to the gate voltages, or equivalently, 
to the random off-set charges. 
The numerical results show that the 
bands are flat to better than one part in  
a thousand, especially at $f=0.48$.  
To understand the underlying 
physics, a tight-binding model is developed.

\subsection{Tight-Binding Model}

\begin{figure}[t] 
\epsfxsize=3.5in 
\epsfbox{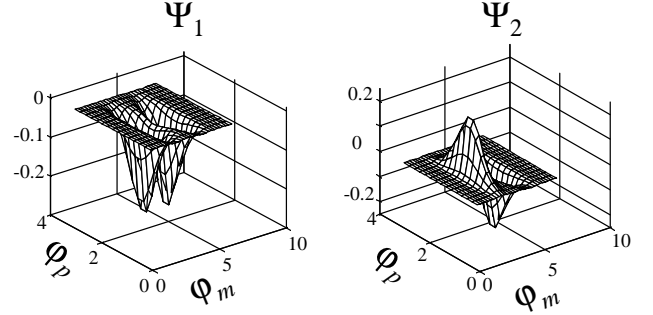} 
         \caption{The eigen wave functions for the lower ($\Psi_1$) 
and upper ($\Psi_2$) energy levels at $f=1/2$  as a function of 
the phases.  
          } 
\label{fig_wavefunction} 
      \end{figure}

Consider the case near the degeneracy point   $f=1/2$. 
The minima in energy 
occurs when $\varphi_p^*=0$ and $\varphi_m=\pm \varphi_m^*$ 
where $\cos\varphi_m^*= 1/2\alpha$. Near the minimum at 
$[\varphi_m,\varphi_p]=[\varphi_m^*, 0]$, the potential looks like 
a double potential well  
repeated at lattice points 
${\bf a}_1=2\pi {\bf i}_x$ and ${\bf a}_2=\pi {\bf i}_x + \pi {\bf i}_y$. 
Fig.~\ref{fig_wavefunction} 
shows the two eigen functions in a unit cell. 
The wave function for the lower level ($\Psi_1$) is symmetric and 
the wave function for the upper level ($\Psi_2$) is antisymmetric. 
Both of the wave functions are   localized  
near the two minima in $U$ in the unit cell.

To find an  approximate tight-binding solution, 
let  $u(\varphi_m,\varphi_p)$ be the wave function for 
the ground state on one side of  the double potential wells, 
and $v(\varphi_m,\varphi_p)$ be the wave function on the other side. 
The tight-binding solution for $H_t$ in Eqn.~\ref{H_t_pm} 
is $\Phi=c_u u + c_v v$ and satisfies  
    \begin{equation} 
     \left( 
      \matrix{H_{uu} & H_{uv} \cr 
              H_{vu} & H_{vv} \cr} 
             \right)  
      \left( 
         \matrix{c_u \cr c_v \cr} 
      \right) 
   = 
     E  
      \left( 
         \matrix{c_u \cr c_v \cr} 
      \right) 
    \label{H_matrix_tb} 
    \end{equation} 
Because the double well is symmetric at $f=1/2$, each wave function has 
the same energy $\epsilon_0$ and so $H_{uu}=H_{vv}=\epsilon_0$ . 
Let $t_1$ be the tunneling matrix element 
between these two minima in the same unit cell and $t_2$ between 
nearest neighbor minima in the  adjacent unit cells. 
Then  
$H_{uv}=H_{vu}^*= -t_1 - t_2 e^{i{\bf k}\cdot {\bf a}_2} 
- t_2 e^{i{\bf k}\cdot ({\bf a}_1 - {\bf a}_2)}$. 
The eigen energy levels are 
$E=\epsilon_0 \mp |H_{uv}|$. 
The effect 
of $t_1$ is to split the degeneracy of the two states so that at 
$k=0$, the energy is $\epsilon_0 \mp (2t_2 + t_1)$  
for the symmetric and 
antisymmetric states respectively.  The effect of $t_2$ is to 
give dispersion in $k$, that is, in gate voltage  
and off-set charges, to the energy levels.   
Because  we want to minimize the gate-voltage (and off-set charge) dependence, 
we seek to  minimize  the tunneling $t_2$ from one unit cell to another. 
Likewise,  we want the two localized states in the two wells to 
interact, so that  we want $t_1$ to be non-zero. This is why the potential 
landscape in Fig.~\ref{fig_pot} was chosen to have $\alpha \approx 0.8$: 
The potential has a much lower barrier between states in the 
double well, but a large barrier between states from one double well 
to the next. 
 
An estimate of  $t_i$ can be obtained from  calculating the action $S_i$ 
between the two minima and using  
$t_i\approx (\hbar \omega_i/2\pi) e^{-S_i/\hbar}$ 
where $\omega_i$ is the attempt frequency of escape in the potential well. 
The action from point $\vec{\varphi}_a$ to $\vec{\varphi}_b$ is 
    \begin{equation} 
    S=\int_{\vec{\varphi_a}}^{\vec{\varphi}_b} 
   \left( 2 M_{nn}(E-U)\right)^{1/2} \, |d\varphi_{n}| 
     \end{equation} 
Here ${\bf n}$ is a unit vector along the path of integration, 
$d\varphi_{n}$ the differential path length, and  
$M_{nn}= {\bf n}^T\cdot {\bf M} \cdot {\bf n}$ is the component of the 
mass tensor along the path direction. In both cases we will approximate 
the energy difference $E-U$ as the difference in the  
potential energy $\Delta U$ from the minima  along the path. 
 
First, consider the calculation of $t_1$, the tunneling 
matrix element  within the unit cell.  
The path of integration is taken  from $(-\varphi_m^*,0)$ to 
$(\varphi_m^*,0)$ along the direction ${\bf n}={\bf i}_x$, so that 
$M_{nn}=M_m$ for this path. The potential energy at the minima 
is $U_{min}=2-1/2\alpha$. 
The difference in the potential energy 
from the minima at $(-\varphi_m^*,0)$ 
along this path is can we written as 
$\Delta U_1=E_J\{2\alpha\left(\cos{\varphi_m}-1/2\alpha\right)^2\}$. 
The action along this path is then 
   \begin{equation} 
       S_1=\int_{-\varphi_m^*}^{\varphi_m^*} 
\left(4 M_m \alpha E_J \right)^{1/2} 
\left(\cos{\varphi_m} -\frac{1}{2\alpha}\right) \, d\varphi_m 
   \end{equation}   
which yields 
   \begin{equation} 
     S_1=\hbar \left[4 \alpha (1+2\alpha +\gamma)E_J/E_c 
             \right]^{1/2} 
\left(\sin{\varphi_m^*} -\frac{1}{2\alpha} {\varphi_m^*}\right) \,. 
    \label{S_1Eqn} 
  \end{equation} 
 
Now consider $t_2$, the tunneling from unit cell to unit cell. 
For example, take the integration to be from 
$(\varphi_m^*,0)$ to one of its nearest neighbor minima at 
$(\pi - \varphi_m^*, \pi)$.  We will take the path of integration 
to be a straight line joining these two points in the $\varphi_m$-$\varphi_p$ 
plane. This path  is not the optimal trajectory, but  
the difference of this straight line path from the optimal 
trajectory is quadratic in the small  deviations of these two paths. 
The straight line path  is described by  
$\varphi_m=\varphi_m^* + \lambda \varphi_p$ where 
$\lambda= (\pi-2\varphi_m^*)/\pi$; it has a direction of 
${\bf n}=\lambda {\bf i}_x + {\bf i}_y$ and a path length 
of $ds=\sqrt{1+\lambda^2} \, d\varphi_p$.  The mass on this direction is 
$M_2=(M_p +\lambda^2 M_m)/(1+\lambda^2)$. 
The difference of the potential energy along this path form the 
minima energy is $\Delta U_2/E_J= 
-2\cos{\varphi_p}\cos{(\varphi_m^* + \lambda \varphi_p)} 
+2\alpha \cos^2{(\varphi_m^*  
+ \lambda \varphi_p)} + 1/2\alpha$. 
The action is then 
    \begin{equation} 
   S_2= \left[2 M_2 E_J(1+\lambda^2) \right]^{1/2} 
  \int_0^\pi  
%\left\{  
%-\cos{\varphi_p}\cos{(\varphi_m^* + \lambda \varphi_p)} 
%+\alpha \cos^2{(\varphi_m^* + \lambda \varphi_p)} + \frac{1}{2\alpha} 
%             \right\} 
\left(\frac{\Delta U_2}{E_J}\right)^{1/2} \, d\varphi_p\,. 
    \end{equation} 
The integrand  is not analytically  integrable, but it  is zero at the  
end points of the integration and is 
well approximated by 
$\sqrt{\Delta U_2/E_J} \approx (1/\sqrt{2\alpha})\cos(\varphi_p-\pi/2)$. 
With this approximation, 
$   S_2= \left(4 M_2 E_J(1+\lambda^2)/\alpha \right)^{1/2}$, which is 
   \begin{equation} 
    S_2=\hbar \sqrt{  
\frac{E_J}{E_c} \left(  \frac{(1+\gamma)(1+\lambda^2)}{\alpha}+ 2\lambda^2 
 \right)}\,. 
    \end{equation} 
 
To compare the tunneling rates we would first need the  
attempt frequencies in the two directions. However, we can  
consider the attempt frequencies to be of the same 
order of magnitude and thus $t_2/t_1 \sim e^{-(S_2-S_1)/\hbar}$. 
For $\alpha=0.8$, we find that 
$S_1/(\hbar\sqrt{E_J/E_c})\approx 0.6$ and 
$S_2/(\hbar\sqrt{E_J/E_c})\approx1.4$. 
For $E_J/E_c\sim 100$, then $t_2/t_1 \sim  10^{-4} \ll 1$. 
We are therefore  able to ignore $t_2$, 
the tunneling from  the unit cell to unit cell. 
This  means that there is little dispersion in the 
energy levels with ${\bf k}$ and consequently, with 
the voltage  or off-set charges. 
In fact, using the action one can show that for 
$\alpha$ smaller than about 0.85, $t_1 > t_2$ 
for $E_J/E_c \approx 80$. 
Throughout the rest of the paper we will choose parameters so that 
the  effects of $t_2$ can be ignored.

We now obtain an approximate solution for the  
energy levels and tunneling matrix elements 
by modeling each side  of the double potential. 
Near the minimum at 
$[\varphi_m,\varphi_p]=[\varphi_m^*, 0]$, the potential looks like 
an anisotropic two-dimensional harmonic oscillator. The  
Hamiltonian in the vicinity of the minimum is approximately, 
(with $Q^I_i=0$) 
    \begin{equation} 
   {\cal H} \approx 
   \frac{1}{2}\frac{ P_p^2}{M_p} 
+\frac{1}{2} M_p \omega_p^2 \varphi_p^2 
+ \frac{1}{2}\frac{ P_m^2}{M_m} 
+\frac{1}{2} M_m \omega_m^2 (\varphi_m -\varphi_m^*)^2 
+U_0 
    \end{equation} 
where 
   \begin{equation} 
\frac{\hbar \omega_p}{E_J}= \sqrt{\frac{4}{\alpha (1+\gamma) (E_J/E_c)}} 
    \end{equation} 
and  
   \begin{equation} 
\frac{\hbar \omega_m}{E_J}=  
\sqrt{\frac{4(4\alpha^2-1)}{\alpha (1+2\alpha +\gamma) (E_J/E_c)}} 
    \end{equation} 
and $U_o = 2 -1/2\alpha$. 
The ground state $\phi_o$ of the single  harmonic well 
has energy $\epsilon_0= \hbar (\omega_p + \omega_m)/2 +U_o$. 
Let's now use this approximation to understand the energy levels, 
first at $f=1/2$ and then near this point.

At $f=1/2$  we expect  
the four lowest energy levels of the two-dimensional 
harmonic oscillator  to be 
with $\omega_m < \omega _p$,  
$E_1= \epsilon_0 -t_1$, 
$E_2= \epsilon_0 +t_1$, 
$E_3= \epsilon_0 -t_1 + \hbar \omega_m$, 
$E_4= \epsilon_0 +t_1 + \hbar \omega_m$. 
Table~\ref{table_half} compares the results 
and we   also list  the small  
anharmonic corrections to the simple harmonic energy levels. 
We have chosen to 
compare $(E_1 + E_2)/2$ and $(E_3 + E_4)/2$ so that  
the tunneling term is absent and a direct comparison with  
the simple harmonic oscillators can be made. 
   \begin{table}[tbh] 
{\small 
    \begin{tabular}{|l|ccccccc|}\hline 
  &  $\hbar \omega_m $ & $\hbar \omega_p$  
  & $E_0$ & $(E_1+E_2)/2$ & $(E_3+ E_4)/2$ \\ \hline  
Harmonic  & 0.193 & 0.247  & 1.60  & 1.79  & 1.84  \\ 
Anharmonic & 0.183 & 0.238  & 1.59  & 1.77  & 1.83  \\ 
Numerical   & 0.154 & 0.226  &  1.58 & 1.74 & 1.81\\ \hline 
\end{tabular} 
} 
     \caption{A comparison of the energy levels with the 
approximate harmonic oscillator levels 
(with harmonic and  anharmonic terms)  
with the numerical calculations.Here,  
$f=1/2$,  $\alpha=0.8$, $\gamma =0.02$, and 
and $E_J/E_c= 80$. Also  $U_o=1.38$ and $U_{\rm bar}=0.225$ 
for the harmonic and anharmonic estimations. 
All the energies are in units of $E_J$. 
      } 
     \label{table_half} 
   \end{table} 
The agreement between this tight-binding approximation and 
the numerical calculations is good. We have also included the  
barrier height from one minimum to the other one in the same unit 
cell.  
 
If we estimate the attempt frequency for $t_1$ as 
$\omega_m$, then we find that  
for the parameters in Table~\ref{table_half}, 
that the action calculation gives 
$t_1= 10^{-4} E_J$. From the full wave functions, we estimate 
$t_1=(E_2-E_1)/2\approx 10^{-3} E_J$. This discrepancy can be made  
smaller by noting that in the calculation of the action, we could 
more accurately integrate from the classical turning points in 
the potential rather than from the minima \cite{Landau}. 
However, for our purposes, the action expression  
will be sufficient 
for qualitative discussions, and we will use the full numerical  
calculations when estimating actual numbers.

So far we have estimated the energy levels and tunneling matrix elements 
when $f=1/2$. As $f$ is decreased from $f=1/2$ the potential $U$ changes 
such that one well becomes higher than the other,  and the barrier height also 
changes. For  the qubit we are mainly interested in the lowest two 
energy states of the system, so we now estimate 
the terms in tight-binding expression of  Eqn.~\ref{H_matrix_tb}. 
By defining the zero of energy as the average of the two lowest energy  
states at $f=1/2$, we find that  
the Hamiltonian for these two states is  
   \begin{equation} 
      H=\left( 
         \matrix{F & - t \cr 
                -  t & -F \cr} 
      \right) 
    \end{equation} 
Here  $F$ is the energy change of each of the wells measured with 
respect to the energy of the wells at the degeneracy point; 
that is, 
$F=(\partial U/\partial f) \delta f$ 
where  $U$ is the potential energy  energy.
Note that since we will be operating the qubit  just below the 
degeneracy point $f= 1/2$, then   $F<0$.
Also, $t=t_1 + \Delta t$, where $t_1$ is the intracell  tunneling matrix 
element calculated at the degeneracy point and $\Delta t$ is the 
change. 
The eigen values are $\lambda_{1,2}= \mp \sqrt{F^2+t^2}$ where
we have explicitly assumed that $F$ is negative and $t$ is positive.
 
The  eigen vectors are  given as the columns in the rotation matrix 
    \begin{equation} 
      D(\theta)=\left( 
          \matrix{ 
                \cos{\theta/2}  &- \sin{\theta/2} \cr 
                \sin{\theta/2} &   \cos{\theta/2} \cr 
                 } 
                \right) 
    \end{equation} 
where $\theta=-\arctan{t/F}$.   
For example, at the degeneracy point, $F=0$, so that $E=\mp t$ and 
the eigen vectors are $(1/\sqrt{2},1/\sqrt{2})^T$ and 
$(-1/\sqrt{2},1/\sqrt{2})^T$. These are just symmetric and antisymmetric 
combinations of the single well wave functions, as expected. 
For $f$ slightly below $1/2$, we have $|F| \gg t$, so $\theta \approx 0$, 
and the energies are $E=\mp \sqrt{F^2 + t^2}\approx \pm F$. The eigen 
vectors are approximately $(1,0)^T$ and $(0,1)^T$, so that the eigen states 
are nearly localized in each well.

It is more convenient to discuss the Hamiltonian and eigen states 
in the rotated coordinate system 
such that $H_D=D^T(\theta)HD(\theta)$. 
In the rotated coordinate system,  the Hamiltonian is diagonal 
with  
   \begin{equation} 
H_D= - \sqrt{F^2 + t^2} \, \sigma_z 
   \label{HD_eqn2} 
     \end{equation} 
and the eigen energies are $E=\pm \sqrt{F^2 + t^2}$ 
and the eigen states are then 
simply spin down $|0>=(1,0)^T$  and spin up $|1>=(0,1)^T$ vectors. 
In other words, no matter what the operating field is, we can always 
go to a diagonal representation; but the rotation matrix 
must be used  to  relate the simple spin up and down vectors to 
the linear combinations of the wave functions in the well.

\section{Manipulation of the Qubit} 
\label{section-man} 
 
As noted above, the flexibility of the design of Josephson 
junction circuits affords a variety of methods for manipulating 
and controlling the state of qubits.  In this section we show 
how the basic qubit circuit can be modified to allow precise 
control of its quantum states. 
To manipulate the states of the qubit, we need 
control over the properties of the qubit.  
For example, control over $f$, the magnetic field,  allows one 
to change the operating point and $F$, the value of  
the energy difference between the two states.
Control over the potential barrier height allows changing  
of the tunneling through  $t$. 
For example, if the operating point of  $F_o$ and $t_o$ are changed  
by $\Delta F$ and $\Delta t$, then 
the  Hamiltonian in the rotated coordinate system is 
  \begin{equation} 
   H_D=-\sqrt{F_o^2 + t_o^2}\, \sigma_z + \Delta H_D 
    \end{equation} 
where with  $\theta_o= -\arctan t_o/F_o$, 
    \begin{equation} 
      \Delta H_D = 
      \Delta F (\cos{\theta}\, \sigma_z  -\sin{\theta} \,\sigma_x) 
  - \Delta t (\sin{\theta}\, \sigma_z + \cos{\theta}\, \sigma_x) \,. 
    \end{equation}

\begin{figure}[tbh] 
%\hspace{3.5in} 
\epsfysize=2.0in 
\epsfbox{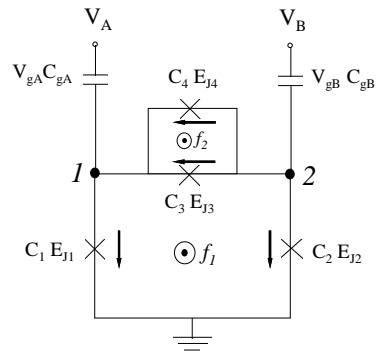} 
     \caption{The four-junction qubit.   
Two junctions form 
a SQUID loop and have Josephson Energies and capacitance $\beta$ times 
larger than the other junctions~1 and 2 which both have 
Josephson energies $E_J$ and capacitance $C$. 
The nodes A and B represent 
the superconducting islands which are coupled by gate capacitors $C_g= 
\gamma C$ to gate voltages $V_A$ and $V_B$. 
The arrows define the direction of the currents.  
The flux is out of  the page.} 
  \label{fig_qubit} 
  \end{figure}

The control over $F$ can be done by changing $f$. 
The control over $t$ can be done by changing the barrier heights. 
To control the barrier heights by external parameters, we replace the 
third junction by a SQUID which acts like a variable strength junction. 
The modified  circuit of the  qubit is shown in Fig.~\ref{fig_qubit}. 
 
We now analyze this circuit since it will be used in all subsequent 
discussion of the qubit. 
Flux quantization around each of the two loops, gives 
$\varphi_1 - \varphi_2 + \varphi_3=- 2\pi f_1$ and 
$\varphi_4- \varphi_3 = -2\pi f_2$. 
The Josephson energy due to each junction is $E_{Jn}(1-\cos\varphi_n)$. 
The total Josephson energy $U$ is then 
   \begin{equation} 
      \frac{U}{E_J}= 2  +2\beta  - 2\cos\varphi_p \cos\varphi_m 
               - 2\beta \cos(\pi f_a)\cos(2\pi f_b + 2\varphi_m) \,, 
   \end{equation} 
where $\varphi_p= (\varphi_1 + \varphi_2)/2$ 
and $\varphi_m= (\varphi_1 - \varphi_2)/2$, and also 
$f_a=f_2$  and $f_b= f_1 + f_2/2$. 
Hence we see that $2\beta \cos(\pi f_a)$ plays the role of 
$\alpha$ in the three junction qubit, but now this term can be  
changed by changing $f_a=f_2$, the flux in the top SQUID loop. 
Likewise, $f_b=f_1+f_2/2$ plays the role of $f$ in the three-junction 
qubit. 
The reduced  Hamiltonian is then 
   \begin{eqnarray} 
H_t= 
 & & \frac{1}{2}\frac{ P_p^2}{M_p}+ \frac{1}{2}\frac{ P_m^2}{M_m} + 
E_J\left\{ 2  +2\beta \right. \nonumber \\ 
 & &\left. - 2\cos\varphi_p \cos\varphi_m 
               - 2\beta \cos(\pi f_a)\cos(2\pi f_b + 2\varphi_m) 
\right\} 
   \end{eqnarray} 
where  
$M_p=(\Phi_o/2\pi)^2 2C(1+\gamma)$ 
and $M_m=(\Phi_o/2\pi)^2 2C(1+4\beta +\gamma)$. 
 
To manipulate the parameters in the Hamiltonian 
let the magnetic fields change very slightly  
away from the  
some degeneracy point of   $f_1^*$ and $f_2^*$ to a new operating point 
$f_1^o= f_1^* + \epsilon_1$ and $f_2^o= f_2^* + \epsilon_2$. 
Then $F$ changes from zero to  $ F_o = r_1 \epsilon_1 + r_2 \epsilon_2$ 
and $t$ changes to  $t_o = t_1 + s_1 \epsilon_1 + s_2 \epsilon_2$, 
where $r_i$ and $s_i$ are constants and $t_1$ is the tunneling  
matrix element at the degeneracy point as found in the previous section. 
We take the operating point to be effectively in the regime where
$f<1/2$ in Fig.~\ref{fig_ebands}, so that 
$\epsilon_{1,2}<0$. Hence, $F_o<0$. Also, $t_o$ is assumed to remain
positive.
In the new  rotated frame with $\theta_o=  - \arctan{t_o/F_o}$, the  
Hamiltonian  given 
by Eqn.~\ref{HD_eqn2} is  
$ H_D =- \sqrt{F_o^2 + t_o^2} \, \sigma_z $. 

Away from this new operating point, let $f_1=f_1^o + \delta_1$ 
and $f_2=f_2^o + \delta_2$.  
In the operation of the qubit, $|\delta_i| \ll| \epsilon_i|$ and 
$\delta_i$ usually will usually  have a sinusoidal time dependence. 
Then  
$ F= F_o + r_1 \delta_1 + r_2 \delta_2$ 
and $t= t_o + s_1 \delta_1 + s_2 \delta_2$, 
so that $\Delta F= r_1 \delta_1 + r_2 \delta_2$ 
and $\Delta t = s_1 \delta_1 + s_2 \delta_2$.
Then  Hamiltonian in the rotated system 
with  $\theta_o =-\arctan{t_o/F_o}$ is
   \begin{equation} 
   H_D= -\sqrt{F_o^2 + t_o^2} \, \sigma_z + \Delta H_D 
    \label{rotated-H}
    \end{equation} 
where 
  \begin{eqnarray} 
    \Delta H_D = & & 
( r_1 \delta_1 + r_2 \delta_2) 
 (\cos{\theta_o}\, \sigma_z  - \sin{\theta_o} \,\sigma_x) 
\nonumber \\ 
 & -& (s_1 \delta_1 + s_2 \delta_2) 
(\sin{\theta_o}\, \sigma_z + \cos{\theta_o}\, \sigma_x) \,.
  \end{eqnarray} 
Hence we see that changes in the magnetic field from 
the operating point of $f_1^o$ and $f_2^o$ 
cause both $\sigma_z$ and $\sigma_x$ types of interactions.

To find  the magnitude of these changes, we calculate 
the coefficients of change ($r_1$, $r_2$, $s_1$ and $s_2$)  
most simply at the degeneracy point where $\epsilon_i=0$; 
that is, at the degeneracy point $f_i^o=f_i^*$. 
We choose the degeneracy  point for the four-junction qubit 
at $f_1^*=1/3$ and $f_2^*=1/3$. This  results in 
classically doubly degenerate levels. In fact, any choice  
which satisfies $2f_1^* +f_2^*=1$ when the classical energy $U$ 
has two minima, will also result in  
doubly degenerate levels. 
For example $f_1^*=1/2$ and $f_2^*=0$ is also a possible and 
convenient choice. However, we prefer $f_1^*=f_2^*=1/3$ for the  
following reason. The change in potential energy with 
$f_a$ gives 
 \begin{eqnarray} 
      \frac{  \partial U}{\partial f_a} &=& 
   -2\pi \beta \sin{\pi f_a} \cos{2 \varphi_m^o}\nonumber \\ 
          \frac{  \partial^2 U }{\partial f_a^2} &=& 
   -2\pi^2 \beta \cos{\pi f_a} \cos{2 \varphi_m^o} 
\end{eqnarray} 
The first order terms vanishes if $f_2^o=0$, resulting in the 
potential barrier always decreasing with changes in $f_2$.  
On the other hand, if $f_2^o=1/3$, then the barrier height can be made 
to increase and decrease with changes in $f_2$, thus allowing more 
control of the qubit. 
 
 Now the coefficients of change  
($r_1$, $r_2$, $s_1$ and $s_2$) can be estimated 
both from the numerical calculations and from the tight-binding model 
as shown in Appendix~B. 
We find that at the degeneracy point of $f_1=f_2=1/3$, 
    \begin{equation} 
         \frac{r_1}{E_J}= 2 \pi \sqrt{1-1/(4 \beta^2)} \,. 
     \end{equation} 
For our  example with  
$\beta=0.8$, we have  $r_1/E_J= 4.90$.  
Estimates  obtained from the numerical calculations 
done by  changing  $f_1$ and 
$f_2$, give $r_1/E_J=4.8$ and $r_2/E_J=2.4$ in good agreement 
with Eqn.~\ref{r_1_eqn} in Appendix~B.

Likewise, from Appendix~B we have that 
$s_1=0$ and 
$s_2=\eta t \sqrt{E_J/E_c}$ 
where $\eta$ is of the order of unity. For the operating point 
we find $\eta \sim 3.5$. Therefore, changes in $H$ due to 
changes in $t_1$ go like $\sigma_x$. 
These tight-binding estimates for  
$\beta=0.8$ give 
$s_1=0$ and $s_2/E_J=0.03$. 
Full numerical calculations for our example  with 
$s_1=0$ and $s_2/E_J=0.20$. The agreement with the tight-binding 
results are good, although the tight-binding underestimates 
$s_2$ with for these parameters. 
 
In summary, 
from  the
degeneracy point of   $f_1^*= f_2^* =1/3$,   let the  operating point be
$f_1^o= f_1^* + \epsilon_1$ and $f_2^o= f_2^* + \epsilon_2$, so that 
$ F_o = r_1( \epsilon_1  \epsilon_2/2)$ 
and  $t_o = t_1 +  s_2 \epsilon_2$.
Now consider the changes in field  about the operating point
such that $f_1= f_1^* + \delta_1$ and $f_2= f_2^* + \delta_2$.
In the rotated frame where  $\theta_o=-\arctan{t_o/F_o}$, 
the Hamiltonian is 
   \begin{equation} 
H_D= -\sqrt{F_o^2 + t_o^2}\, \sigma_z + \Delta H_D 
   \label{fullA_Ham} 
\end{equation} 
where 
  \begin{eqnarray} 
    \Delta H_D = & & 
 r_1 (\delta_1 + \frac{\delta_2}{2}) 
 (\cos{\theta_o}\, \sigma_z  - \sin{\theta_o} \,\sigma_x) 
\nonumber \\ 
 & -&  s_2 \delta_2
(\sin{\theta_o}\, \sigma_z + \cos{\theta_o}\, \sigma_x) \,.
    \label{fullA_Ham2} 
 \end{eqnarray} 
and  $r_1/E_J=2 \pi \sqrt{1-1/(4 \beta^2)} $ and 
$s_2=\eta t_o \sqrt{E_J/E_c}$. 
 
A typical design for a qubit will have 
$E_J/E_c=80$, $\beta=0.8$, $\gamma=0.02$. We find from numerical
calculations
that $t_o \approx 0.005 E_J$ and $\eta\approx 3.5$, which agree well
with our tight-binding estimates.
We operate at $f_1=f_2=0.33$ so that $\epsilon_1=\epsilon_2=
-1/300$. (This is equivalent to operating the three-junction
qubit at $f=f_1 + f_2/2= 0.495$ in Fig.~\ref{fig_ebands}.)
Writing the energies as   $E_i= h \nu_i$,
we have taken typical values of $E_J= 200\,{\rm GHz}$
and $E_c= 2.5\,{\rm GHz}$, and we find that 
$t_o= 1\,{\rm GHz}$ and $F_o= 5\,{\rm GHz}$ (which gives
a splitting between the two states of about  $10\,{\rm GHz}$).
The Hamiltonian is found to be 
   \begin{equation} 
\frac{H_D}{E_J}= - 0.025 \sigma_z
+ 
\left(4.0 \delta_1 + 2.1 \delta_2 \right)\sigma_z 
- \left(0.46 \delta_1 + 0.41  \delta_2 \right)\sigma_x 
\label{DHDnumbers} 
    \end{equation} 
The numerical values used are from numerical calculations. 
These values agree well with the estimates used in 
Eqns.~\ref{fullA_Ham}  and~\ref{fullA_Ham2} for the level splitting
and the terms proportional to $r_1$; the terms proportional
to  $s_2$ match to about 50\%, due to the more sensitive nature 
of estimating the tunneling terms.

The terms containing $\sigma_x$ can be used to produce
Rabi oscillations between the two states by modulating
$\delta_1$ and $\delta_2$ with microwave pulses of 
the frequency of the level splitting of $2F_o = 10\,{\rm GHz}$.
One could arrange the values of $\delta_1$ and $\delta_2$ to
make the time-varying $\sigma_z$-term vanish. Then the operation of 
the qubit would be isomorphic to the NMR qubit. However,
our simulations show that such an arrangement couples 
higher energy levels and invalidates the simple two-state 
approximation. This is due to the large matrix element
between the ground state and the second excited state
given by the change in potential due to  varying $\delta_2$.
(It is interesting to note that similar coupling to higher levels
occurs in qubits based on  the RF SQUID  and on simple
charge states.) 
We propose to manipulate the qubit by  varying $\delta_1$ which
causes a Rabi oscillation through the $\sigma_x$ term as well
as a strong modulation of the Larmor precession through the
time varying $\sigma_z$ term. Because the Rabi frequency is much 
smaller than the Larmor frequency, the precession causes
no problem for manipulating the qubit.  For $\delta_1=0.001$ and
$\delta_2=0$, the
Rabi frequency is about  90\thinspace MHz. We note that this mode of
operation is also possible with the three-junction qubit.
Of course, it will not be possible to completely eliminate
the  deleterious effects of the $\delta_2$ coupling, but
the  effect of this coupling can be greatly  reduced
if $\delta_2$ is restricted below 0.0001.

The varying magnetic fields $\delta_1$ and $\delta_2$ can
be applied locally to the qubit by using  a control 
line to inductively  couple to the qubit. Moreover, if the
the control line is driven by an Josephson oscillator, then
the coupling circuit could be fabricated on the same chip.

\section{Interaction between Qubits}

\begin{figure} 
\epsfxsize=3.5in 
\epsfbox{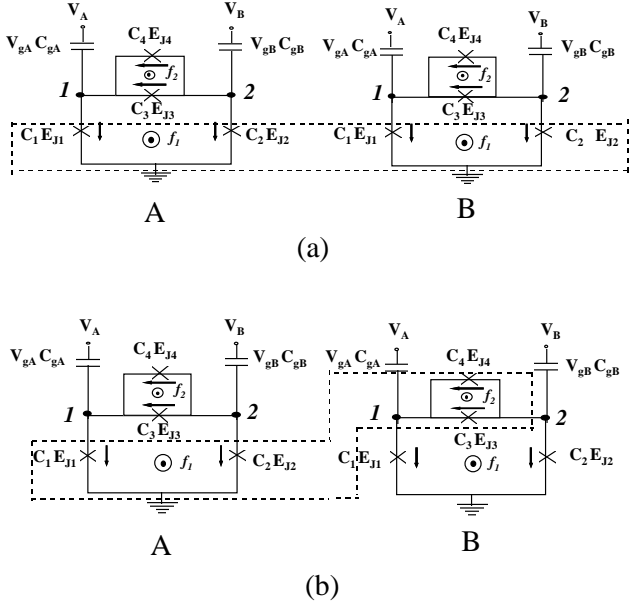} 
        \caption{ Coupling of qubits $A$ and $B$  through the 
mutual inductance between (a)~the lower regions of both, 
and (b)~the lower region of $A$ and the upper region of $B$.} 
 \label{coupling_fig} 
  \end{figure}

A variety of methods are available for coupling qubits together. 
As noted in \cite{Lloyd95b,Deutsch95}, essentially any interaction 
between qubits, combined with the ability to manipulate qubits 
individually, suffices to construct  
a universal quantum logic gate. 
Here we present  two methods for coupling qubits inductively
as shown in Fig.~\ref{coupling_fig}. 
The inductive coupling could either be permanent, or  
could be turned on and off at will by inserting Josephson 
junctions in  the coupling loops. 
 
Fig.~\ref{coupling_fig}a shows  
one way of coupling  two identical qubits. 
The lower portions of each qubit (the loops that contain the 
circulating currents) are inductively  coupled. 
To a first 
approximation we model the coupling as changing the flux in  
each of the two lower rings only through the mutual inductive 
coupling. (We ignore the the self-inductance 
which can easily be included.) 
The effective frustration in the lower loop of  $A$, $\tilde{f}^A_1$, is 
changed over the applied frustration $f^A_1$ to 
$\tilde{f}^A_1= f^A_1 + M I^B_1/\Phi_o$. Here the  
current in the lower loop of $B$ is $I^B_1$. 
Similarly, $\tilde{f}^B_1= f^B_1 + M I^A_1/\Phi_o$.  
The coupled Hamiltonian is  
  \begin{equation} 
  H_{AB}= H^A(\tilde{f^A_1}) + H^B(\tilde{f^B_1}) + M I^A_1 I^B_1 
   \end{equation} 
which is  
the sum of the Hamiltonians for each system plus 
a  term due to the mutual inductive coupling. 
 
The inductively coupled contribution to 
the frustration is estimated to be of the order of  
$10^{-3} \Phi_o$ which is much smaller than the applied frustration. 
Since each persistent current will inductively couple into
the other qubit, this will produce changes in the Hamiltonian
of the $\sigma_z$ and $\sigma_x$ type and these changes will
be proportional to the sign of the circulating currents in
the qubit.
Hence, we expect the coupling to be described by  an
interaction  Hamiltonian of the form,
   \begin{equation} 
  H_{AB}^{\rm int} = \kappa_1 \, \sigma_z^A \sigma_z^B 
  + \kappa_2 \sigma_z^A \sigma_x^B  + + \kappa_3 \sigma_x^A \sigma_z^B  
   \label{Hint2} 
    \end{equation} 
Hence we see that this interaction has both  
a $\sigma_z^A \sigma_z^B$ and a  
$\sigma_z^A \sigma_x^B$ types of coupling. 
We have estimated magnitude of $\kappa_i\approx
0.01 E_J$.

As  Eqn.~\ref{DHDnumbers} shows,  
the inductive coupling between the 
qubits can be made to be a substantial fraction of the  
qubit Larmor frequency.  This is an attractive feature, 
as the coupling between two qubits sets the speed limit 
for how rapidly two qubit quantum logic operations 
can be performed in principle.  In practice, it may 
be desirable to sacrifice speed of operation for 
enhanced accuracy: in this case, the inductive coupling 
could be designed to be smaller by decreasing the overlap of the  
inductive loops with the circuits.

Coupling between qubits is similar to the coupling we envision between
the qubit and the measurement circuits containing SQUID-like
detectors.  In its usual configuration, the SQUID is biased  in the
voltage state which produces a voltage related to the flux through its
detector loop.  However, such a strong, continuous measurement on a
qubit would destroy the superposition of states in the qubit and
project out only one of the states.  This problem can be circumvented
by designing a SQUID such that it is current biased in the
superconducting state and hence is not measuring the flux in its
detector loop.  When one needs to measure the qubit, the SQUID can be
switched to its voltage state, for example, by applying a pulse of
bias.  The coupling from mutual inductance between the SQUID and the
qubit will also have to be controlled. Other measurement schemes
using SQUIDs which are weakly coupled to the macroscopically
coherent system have been proposed
\cite{Tesche}.

\section{Computing with the PC Qubit} 
 
All the ingredients for quantum computation are now available. 
We have qubits that can be addressed, manipulated,  
coupled to each other, and read out.  As will be indicated below, 
the particular qubits that we have chosen are well insulated 
from their environment as well.  The flexibility of design 
for collections of qubits now allows a wide variety of overall 
designs for quantum computers constructed from such qubits. 
 
Before discussing various superconducting quantum computer 
architectures, let us review some basic ideas about quantum  
logic and see how to implement quantum logic using our 
superconducting qubits.  A quantum logic gate is a unitary 
operation on one or more qubits.  Quantum computations are 
typically accomplished by building up quantum logic circuits 
out of many quantum logic gates.  Just as in the case of 
classical computers, certain sets of quantum logic 
gates are universal in the sense that any quantum computation 
can be performed by wiring together members of the set. 
In fact, almost any interaction between two or more qubits 
is universal \cite{Lloyd95b,Deutsch95}; but a convenient universal 
set of quantum logic gates widely used in the design of 
quantum algorithms consists of single qubit rotations 
and the quantum controlled-NOT gate, or CNOT \cite{Barenco95}.     
 
\subsection{One-Qubit Rotation} 
 
An arbitrary one qubit rotations can be written as  
$e^{-i\sigma t} = \cos t -i \sin{t} \, \sigma$ for some Pauli matrix  
$\sigma = a \sigma_x + b\sigma_y + c \sigma_z$, where  
$a^2+b^2+c^2=1.$  There are many ways of accomplishing a 
one qubit rotation: the ability to rotate the qubit 
by a precise amount around any two orthogonal axes 
suffices.   Pursuing the analog with NMR, we choose a method 
that involves applying an oscillatory field applied at 
the qubit's resonant frequency to rotate the qubit.

The Hamiltonian for a single qubit ($A$) can be gotten from  
Eqn.~\ref{DHDnumbers}. Here we assume $E_J= 200\,{\rm GHz}$,
$\delta_1= 0.001 \cos{\omega t}$
and $\delta_2=0$, and the level splitting is $ \omega =10\,{\rm GHz}$.
Then, the Hamiltonian is
   \begin{equation} 
H_D({\rm GHz}) = 5\, \sigma_z
+ 
0.80\, (\cos{\omega t}) \, \sigma_z 
-0.09\,  ( \cos{\omega t})\, \sigma_x 
\label{DHD2numbers} 
    \end{equation} 
The Rabi frequency is $90\, MHz$ so that a $\pi$ pulse would
be about 20\thinspace nsec.

\subsection{Two-Qubit Controlled NOT} 
 
A controlled NOT is a two qubit quantum logic gate that flips 
the value of the second qubit iff the value of the first qubit 
is 1.  That is, it takes: $|00\rangle \rightarrow |00\rangle$, 
$|01\rangle \rightarrow |01\rangle$, $|10\rangle \rightarrow 
|11\rangle$, and $|11\rangle\rightarrow |10\rangle$. 
A controlled NOT can be combined with single qubit rotations 
to give arbitrary quantum logic operations.  A controlled NOT 
can be straightforwardly implemented in the superconducting 
qubit system by exploiting the analogy with NMR.   
Suppose that two qubits $A$ and $B$ have been constructed 
with an inductive coupling between their lower loops 
as in the first part of the previous section.
Then the level splitting of qubit $B$ depends on the
state of qubit $A$, with values $\Delta E_{A,0}$ for
$A$  in the $|0>$ state and 
$\Delta E_{A,1}$ for $A$  in the $|1>$ state.
When a resonant pulse corresponding of 
$\Delta E_{A,1}/\hbar$ is applied to qubit $B$, it will only 
change if qubit $A$ is in its $|1>$ state.
Since the coupling between the qubits is 
considerably larger than the Rabi frequency, the amount of time  
that it takes to perform the controlled NOT operation is 
equal to the amount of time it takes to perform a $\pi$ rotation 
of a single qubit. 
 
So the basic quantum logic operations can be performed on our 
superconducting qubits in a straightforward fashion.  Accordingly, 
it is possible in principle to wire groups of qubits together to construct 
a quantum computer.  A variety of architectures 
for quantum computers exist, usually consisting of regular 
arrays of quantum systems that can be made to interact either 
with their neighbors or with a quantum ``bus'' such as a cavity 
photon field or a phonon field in an ion trap that communicates 
equally with all the systems in the array.   
Because of the flexibility  
inherent in laying out the integrated Josephson junction circuit, 
a wide variety of architectures are possible.  A particularly 
simple architecture for a quantum computer can be based on the 
proposal of Lloyd\cite{Lloyd93,Lloyd_SciAm95}  
for arrays of quantum systems such 
as spins or quantum dots.

\subsection{Linear Chain of Qubits} 
 
Consider a linear array of qubits $ABABABAB \ldots$.  Let the 
bottom of each qubit be inductively coupled to the top of the 
neighbor to the left.  Also let each type of qubit, $A$ and $B$,  
have a slightly different Josephson energy. 
Each qubit also has the  
area of the top loop which is half that of the bottom loop. 
In the absence of the driving electromagnetic fluxes (the $\delta_{i}^j$), the Hamiltonian for 
the system can be  generalized to be written as 
   \begin{equation} 
      H= -\hbar \sum_k \left( 
                     \omega_k \sigma_k^z + 2 J_{k,k+1} \sigma_k^z \sigma_{k+1}^z 
                        \right) 
    \end{equation} 
where $\hbar \omega_k = \sqrt{F_k^2 +t_k^2}$ and $J_{k,k+1}= \kappa_{k,k+1}(r_{1,k} + r_{1,k+1})/2$. 
This problem then maps on the linear chain of nuclear spins which was shown by Lloyd 
\cite{Lloyd_SciAm95} to be a universal quantum computer. 
The coupling needed to perform $\pi/2$ pulses is provided by the  
terms containing the $\delta_{i}^j$'s.  
The nice feature of this linear chain is that separate control lines for AC fields are not needed. 
The whole linear array can sit in a microwave  cavity and be pulsed at the desired frequency. 
(The dc bias fields to ensure $f_1=f_2=1/3$ will require at least two dc control lines). 
The frequencies needed are around 10---25\thinspace GHz with intervals of 
1\thinspace GHz (and  with resolution of about 0.1\thinspace GHz). 
We could make these numbers larger or smaller if needed. 
 
Details of computing with this are given in various references, see for examples Ref.~\cite{Lloyd_SciAm95} 
and Chapter~20 of Ref.~\cite{Bermanbook}.

\subsection{Superconducting Quantum Integrated Circuits} 
 
\begin{figure}[t] 
\epsfxsize=3.5in 
\epsfbox{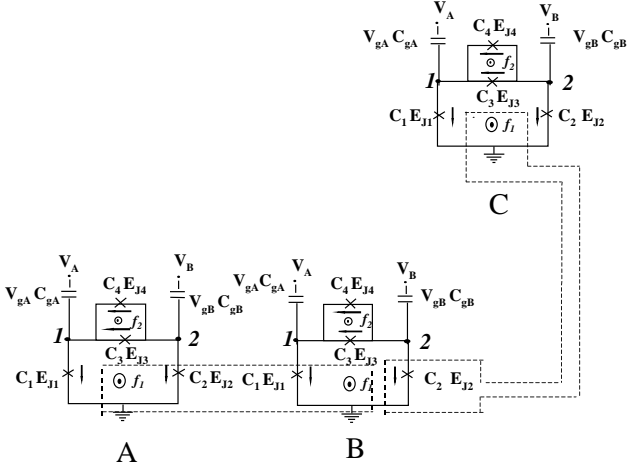} 
  \caption{ 
   A method for coupling a single qubit to other qubits. 
          } 
   \label{fig-sc-ics} 
   \end{figure}

There is no reason why the inductive loops cannot couple 
qubits that are far apart.  In addition, a single qubit can be 
coupled to several other qubits as shown in Fig.~\ref{fig-sc-ics}. 
This arrangement requires separate AC control lines for 
each of the qubits, which then demands localized on-chip 
oscillators.  One can build up essentially arbitrary integrated  
circuits of superconducting qubits by this method.   
This flexibility 
in the construction of quantum computer architectures is one 
of the benefits of using superconducting Josephson junction 
circuits to perform quantum computation.  The quantum 
integrated circuit could be set up to provide a number 
of useful features.  For example \cite{Cahn}, one might be able to 
design the circuit and interactions in such a way that  
it automatically implements an intrinsically fault-tolerant 
quantum computer architecture such as those proposed 
by Kitaev\cite{Kitaev97} and Preskill\cite{Preskill}.   
In addition, 
since the circuits are parallelizable in that different 
quantum logic operations can be performed in different 
places simultaneously, the circuit could be designed to 
provide the maximum possible parallelization of a particular   
problem such as factoring\cite{Shor},  
database search\cite{Grover}, 
or computing a discrete quantum Fourier  
transform\cite{Shor,Simon}.

\section{Decoherence} 
 
We have shown how superconducting circuits can be used to  
construct qubits and quantum logic circuits.   
These superconducting 
qubits have been idealized in the sense that we have ignored 
the effects of manufacturing variability, noise and decoherence. 
Manufacturing variability can be compensated  
for as discussed above: 
before performing any quantum computations, the properties 
of individual qubits can be measured, recorded in a look-up 
table in a conventional computer, and used either 
to supply compensating calibration fields or  
to alter the frequencies 
with which control pulses are supplied to the qubits.   
 
From the point of view of the ultimate performance of a 
superconducting computer, a more pressing issue is that of 
environmentally induced noise and decoherence.  In real systems 
the performance of a qubit will be limited by dissipative 
mechanisms which cause the quantum state to decohere in time 
$\tau_d$.  The `quality factor' for a qubit is the decoherence 
time divided by the amount of time it takes to perform fundamental 
quantum logic operations\cite{DiVincenzo95}.   
The quality factor gives the  
number of quantum logic operations that can be  
performed before the  
computation decoheres, and should be $10^4$ or greater for the  
quantum computer to be able to perform arbitrarily long quantum 
computations by the use of error-correction  
techniques\cite{Shor96,Steane96,Shor96a,Cirac96,Knill97}.

 Decoherence can be due to ``internal''  dissipation 
(quasiparticle resistance), 
or coupling to an environmental degree of freedom. 
It is also possible to couple to an environmental  
degree of freedom, without a dissipative mechanism, that 
will still lead to decoherence\cite{Stern90} 
 
We will now discuss some of the major sources of decoherence.
 
Normal state quasi-particles can 
cause dissipation and energy relaxation at finite 
temperatures in Josephson junctions. However, 
mesoscopic aluminum junctions have been shown to  
have the BCS temperature dependence for the density 
of quasi-particles. At low temperatures this density 
is exponentially small\cite{Tesche90}, so quasi-particle tunneling 
will be strongly suppressed at low temperatures  
and at low voltages, as was seen in a system with  
multiple superconducting islands in Ref.~\cite{vdWal98}. 
We estimate a lower bound of 
$10^4$ for the quality factor, given a sub-gap resistance
of $10^{10}\,\Omega$\cite{Tesche90}.

The qubit can also decohere by emitting  photons.
We estimate this by classically estimating the rate photons
are emitted  by magnetic dipole radiation from oscillating current in the 
loop defining the qubit.  For a loop with radius $R$ with an alternating 
current of frequency $\nu$ and rms amplitude of $I_m$,
the power transmitted to free space is
$P_m=K_mR^4I_m^2\nu^4$ where $K_m=8\mu_o\pi^5/3c^3$.
A typical rate for photon emission is  $1/t_m=P_m/h\nu$, which gives 
an estimate of the decoherence time of 
$t_m=h/K_mR^4I_m^2\nu^3$. An estimate for 
the  frequency is the Larmor frequency,
(other characteristic frequencies such as the Rabi frequency are even smaller).
For our qubit $R\approx 1\,\mu{\rm m}$, $\nu \approx 10\, {\rm GHz}$ and
$I_m=1$\thinspace nA, we find that $t_m \sim 10^7\,{\rm s}$, so that this
is not a serious source of decoherence. However, it should be noted that 
proposals for using RF SQUIDs for qubits,  involve currents of the order of
1\thinspace $\mu$A and and loops of the order of 10\thinspace $\mu$m.
These RF SQUIDs have
$t_m \approx   10^{-3}\, {\rm sec}$, which is substantially 
lower than for our qubits which can be made much smaller have much 
less current.

Inhomogeneity in the magnetic flux distribution can
also be a source of decoherence. This is similar
to $T_2$ in NMR systems. We estimate this for our
system by calculating the amount of flux a 
$1\, \mu{\rm m} \times 1\, \mu{\rm m}$
wire carrying 100\thinspace nA of current induces
in a loop of the same size which has its center 3\thinspace $\mu$m
away. We find that the induced frustration is about $\delta f = 10^{-7}$.
If this is taken as an estimate of the typical variance of the
frustration that difference qubits experience, then
there will be a spread of operating frequencies among the loops.
An estimate of $t_d$ is the time for the extremes of this frequency differ
by $\pi$.  This results in $t_d\approx \pi/(2 r_1\delta f)$,
where we have taken the larger value from Eqn.~\ref{DHDnumbers}.
With $r_1/\hbar \approx 600 \, {\rm GHz}$, we find
$t_d\approx 1.5 \, {\rm msec}$. The dipole-dipole
interaction between qubits gives a time of the same order.

We have also estimated the magnetic coupling between the 
dipole moment of the current loops and the magnetic moments
of the aluminum nuclei in the wire. At low temperatures where
the quasi-particles are frozen out, the decoherence time for a single
qubit is 
of the order of $T_1$ which is exponentially large in the
low-temperature superconducting state.
For an ensemble of qubits, the 
decoherence time time may be of the order of
milliseconds due to the different configurations 
of nuclear spins in the different qubits. 
However, this effect may be reduced  by aligning the spins
or by applying compensating pulse sequences.

Coupling to ohmic dissipation in the environment has 
been modeled for  superconducting qubits operating 
in the charging regime\cite{Shnirman97}. They found 
that source of decoherence could be made sufficiently  
small such that the quality factor is large enough. 
Similar calculations for qubits in the superconducting 
regime of circulating currents have not yet  been done. 
Experiments to measure this decoherence time in 
our circuits are underway. 
In practice electromagnetic coupling to the normal 
state ground plane can 
limit coherence\cite{Rouse95}; however, a superconducting 
ground plane can greatly reduce this coupling.

Other possible sources of  
decoherence are the effects of the measuring circuit, 
the arrangement and stability 
of the control lines for the magnetic fields, 
the ac dielectric losses in the substrate at microwave 
frequencies. These and other source of decoherence will 
have to be estimated in a real circuit environment and 
measured.

Taking  $0.1\, {\rm msec}$ as a lower bound on the 
decoherence time and $10\, {\rm nsec}$ as a switching 
time, we find that the quality factor is of the order 
of $10^4$. Furthermore, if the proper set of 
topological excitations is used to store 
information, the  decoherence time for quantum 
computation can be made substantially longer than 
the minimum decoherence time for an individual 
junction circuit\cite{Kitaev97}.

\section{Summary}

In this paper we have discussed  
a superconducting qubit which  
has circulating currents of opposite 
sign as its two logic states.  
The circuit consist of 
three nano-scale Josephson junctions connected in 
a superconducting loop and controlled by magnetic fields. 
One of the three junctions is a variable junction  
made as a SQUID loop. 
This qubit has quantum states which 
are analogous to a particle with  an anisotropic mass 
moving in an two-dimensional periodic potential. 
Numerical calculations of the quantum states of 
the qubit have been made as well as physical  
estimates  from a tight-binding approximation. 
The advantages of this qubit is that it can be  
made insensitive to background charges in the substrate, 
the flux in the two states can be detected, and 
the states can be manipulated with magnetic fields.  
Coupled systems of qubits are also discussed as well 
as sources of decoherence. 
 
\section*{Acknowledgment} 
This work is supported by 
ARO grant DAAG55-98-1-0369.
TPO acknowledges  support from the Technical University 
of Delft and the Center for Superconductivity at the 
University of Maryland where part of this work was done. 
JJM is supported by the Fulbright Fellowship and from
DGES (PB95-0797).
The work in Delft is financially supported by the Dutch
Foundation for Fundamental Research on Matter (FOM).
We thank Kostya Likharev, Kees Harmans,  Peter Hadley, Chris Lobb,
Fred Wellstood, and Ton Wallast for stimulating and useful discussions.

\section*{Appendix A: Classical Stability} 
 
In this appendix we find the eigen values of the stability 
matrix for the three junction potential and  
the range of frustration around $f=1/2$ where  
there are two stable classical solutions with opposite 
circulating currents. 
 
The potential energy of the Josephson energy of the   
three junction qubit 
is given by Eqn.~\ref{threejun_U} 
 
\begin{equation} 
\tilde{U} =\frac{U}{E_J}= 
2 + \alpha - \cos \varphi_1 - \cos \varphi_2 
		- \alpha \cos (2\pi f + \varphi_1 - \varphi_2) 
\end{equation} 
We are interested in minimum energy phase configurations; that is, stable 
solutions of the following system of equations: 
\begin{eqnarray} 
\frac{\partial \tilde{U}}{\partial \varphi_1} & = &  
\sin \varphi_1 + \alpha \sin (2\pi f + \varphi_1 - \varphi_2) = 0 \nonumber \\ 
\frac{\partial \tilde{U}}{\partial \varphi_2} & = &  
\sin \varphi_2 - \alpha \sin (2\pi f + \varphi_1 - \varphi_2) = 0  
\end{eqnarray} 
The solutions ($\varphi_1^*,\varphi_2^*$) complies with: 
$\sin \varphi_1^*=-\sin \varphi_2^*=\sin \varphi^*$. Then 
\begin{equation} 
\sin \varphi^* = -\alpha \sin (2\pi f + 2 \varphi^*)  
\label{phi*} 
\end{equation} 
 
In order to check the character of the solution we  
compute the 
eigen values of the stability matrix, 
$\frac{\partial^2 \tilde{U}}{\partial 
 \varphi_i \partial \varphi_j}$, where 
\begin{eqnarray} 
\frac{\partial^2 \tilde{U}}{\partial \varphi_1^2} & = &  
\cos \varphi_1 + \alpha \cos (2\pi f + \varphi_1 - \varphi_2) \nonumber \\ 
\frac{\partial^2 \tilde{U}}{\partial \varphi_2^2} & = &  
\cos \varphi_2 + \alpha \cos (2\pi f + \varphi_1 - \varphi_2) \\ 
\frac{\partial^2 \tilde{U}}{\partial \varphi_1 \partial \varphi_2} & = &  
- \alpha \cos (2\pi f + \varphi_1 - \varphi_2) \nonumber 
\end{eqnarray} 
For the states with $\cos \varphi_1^*=\cos \varphi_2^*=\cos \varphi^*$
(these are the ones we are interested here), the eigen values are
\begin{eqnarray} 
\lambda_1 & = & \cos \varphi^* \nonumber \\ 
\lambda_2 & = & \cos \varphi^* + 2\alpha \cos (2\pi f + 2\varphi^*) 
\end{eqnarray} 
 
When $f \neq 0, 1/2$ we have used relaxation  
methods for 
computing $\varphi^*$.  
Both eigen values are greater than 
zero which assures the  minimum  
energy condition. Fig.~\ref{fig_Eclass} 
shows the energy of the minimum energy 
configurations for  $\alpha=0.8$. 
We find that there exists a  
region of values of the field for which two 
different minimum energy phase configuration coexist. 
 
Next we  calculate 
the critical values of the external 
field for this coexistence. We can restrict our analysis to the region 
around $f=0.5$; that is, $[0.5-f_c,0.5+f_c]$ (where $f_c \ge 0$). 
These extrema values of the field correspond to solutions for which one 
of the eigen values is positive and the other equals zero. The inset of 
Fig.~\ref{fig_Eclass} shows $f_c(\alpha)$. 
 
We first calculate $f_c$ when  $\alpha \geq 1.0$. 
The first eigen value which equals zero is $\lambda_1$. Then at $f=0.5 \pm f_c$, 
$\lambda_1=0$ which implies  
$\varphi^*=\mp \pi/2$ mod $2\pi$ (here and below we associate the sign in 
$f_c$ with the sign of the phase in  
order to have $f_c \ge 0$). Then, going to Eqn. \ref{phi*} we get 
\begin{eqnarray} 
\sin (\mp \pi/2) &=& - \alpha \sin (\pi \pm 2\pi f_c \mp \pi) \nonumber \\ 
\pm 1 &=& \pm \alpha \sin (2\pi f_c) 
\end{eqnarray} 
and  
\begin{equation} 
f_c=\frac{1}{2\pi}\arcsin \frac{1}{\alpha} \,. 
\end{equation}

We now calculate $f_c$ when  $0.5 \leq \alpha \leq 1.0$. 
Now the first eigen value to equal  zero is $\lambda_2$, 
and we have to solve: 
\begin{eqnarray} 
\sin \varphi^* &=& - \alpha \sin (2\pi f + 2 \varphi^*) =  
\alpha \sin (\pm 2\pi f_c + 2 \varphi^*) 
\nonumber \\ 
\cos \varphi^* &=& - 2 \alpha \cos (2\pi f + 2 \varphi^*)  
= 2 \alpha \cos (\pm 2\pi f_c + 2 \varphi^*)  
\end{eqnarray} 
We will use $\Delta=\pm 2\pi f + 2 \varphi^*$, so that  
\begin{eqnarray} 
1&=&\sin^2 \varphi^* + \cos^2 \varphi^* \nonumber\\ 
 &=&  \alpha^2 \sin^2 \Delta + 
4\alpha^2 \cos^2 \Delta \nonumber \\ 
& = & \alpha^2 + 3 \alpha^2 \cos^2 \Delta 
\end{eqnarray} 
Then  
\begin{eqnarray} 
\cos \Delta = \sqrt{ \frac{1-\alpha^2}{3\alpha^2} } 
\; \; &;& \; \; \Delta= \mp \arccos(\sqrt{ \frac{1-\alpha^2}{3\alpha^2} })  
\nonumber \\ 
\cos \varphi^* = 2 \sqrt{ \frac{1-\alpha^2}{3} } 
\; \; &;& \; \; \varphi^* = \mp \arccos(2 \sqrt{ \frac{1-\alpha^2}{3} }) 
\end{eqnarray} 
Here we have followed the solution corresponding to 
$\cos (\varphi^*) \ge 0$. Finally we have the solution for 
$f_c$ ($\Delta=\pm 2\pi f_c + 2 \varphi^*$) 
 
\begin{equation} 
f_c= \frac{1}{2\pi} \left[ 2 \arccos( 2 \sqrt{ \frac{1-\alpha^2}{3} }) 
-\arccos( \sqrt{ \frac{1-\alpha^2}{3\alpha^2} }) 
\right] 
\end{equation}

\section*{Appendix B: Tight-Binding Estimate of Coefficients of Change} 
 
Recall that  $f_a=f_2$ and $f_b=f_1+f_2/2$. Assume that we change  
$f_a$ and $f_b$ independently.   
The minima  in $U$ occur at   
$\varphi_p^*=0$ and $\varphi_m^*=\pm \varphi^o_m$ 
Therefore, the energy due to the potential energy is for each of the  
minimum 
   \begin{equation} 
      \frac{U}{E_J}= 2  +2\beta  - 2 \cos\varphi_m^* 
               - 2\beta \cos(\pi f_a)\cos(2\pi f_b +  2\varphi_m^*) \,. 
   \end{equation} 
The change in the   magnetic flux $f_a$ by $\delta f_a$ causes a change 
in $U$ of 
   \begin{equation} 
      \frac{  \partial U}{\partial f_a} \, \delta f_a = 
   -2\pi \beta \sin{\pi f_a} \cos{2 \varphi_m^o} \, \delta f_a  
   \end{equation} 
which  is the same for the minimum at $\pm \varphi_m^o$. 
Whereas, the flux $f_b$ causes a change 
   \begin{equation} 
      \frac{  \partial U}{\partial f_b} \, \delta f_b = 
   \mp 4\pi \beta \cos{\pi f_a} \sin{2 \varphi_m^o} \, \delta f_b 
   \end{equation} 
which has opposite signs for the two minimum. 
Therefore, 
   \begin{equation} 
\frac{\Delta U}{E_J} =     -2\pi \beta \sin{\pi f_a} \cos{2 \varphi_m^o}\, \delta f_a 
\,{\bf 1} 
    - 4\pi \beta \cos{\pi f_a} \sin{2 \varphi_m^o}\, \delta f_b\, 
\sigma_z 
    \label{DeltaU_eqn} 
   \end{equation} 
 
Recall that $\Delta F$  is the change is the  energy  between the  
two states when there is no tunneling. This is  the second 
term in Eqn.~\ref{DeltaU_eqn}, since the first term is only a constant 
for both levels, so that 
   \begin{equation} 
\frac{\Delta F}{E_J} =     
    - 4\pi \beta \cos{\pi f_a} \sin{2 \varphi_m^o}\, \delta f_b\, 
\sigma_z 
    \label{DeltaF_eqn} 
   \end{equation} 
For this change 
$\Delta F= r_1 \delta_1 + r_2 \delta_2$; and since 
$\delta f_b= \delta_1 +\delta_2/2$, we have 
$r_1= 2r_2$ and 
     \begin{equation} 
\frac{r_1}{E_J}=4\pi \beta \cos{\pi f_a} \sin{2 \varphi_m^o}\,. 
   \label{r_1_eqn} 
   \end{equation} 
We have found previously that 
$\cos{ \varphi_m^o}= 1/2\alpha$ where $\alpha=2 \beta \cos{\pi f_a}$ 
so that with $f_a=1/3$, 
    \begin{equation} 
         \frac{r_1}{E_J}= 2 \pi \sqrt{1-1/(4 \beta^2)} \,. 
     \end{equation}

To find the changes in $\Delta t$, we see that the 
changes in 
$t_1=(\hbar \omega_m /2\pi) e^{-S_1/\hbar}$ are dominated by 
changes in $S_1$, so that 
\begin{equation} 
\Delta t= -\frac{t}{\hbar} 
\sum_{i=a,b}\frac{\partial S_1}{\partial f_i} \delta f_i\,. 
\label{deltat_S} 
\end{equation} 
The changes in $f_b$ do not change $S_1$ to first order. 
Hence, changes in $t$ come from changes in $f_a=f_2$ only, so that $s_1=0$. 
But changes in $f_a$ are equivalent to changes in 
$\alpha$ in the three junction problem, so we can use 
Eqns.~\ref{deltat_S} and~\ref{S_1Eqn} and the fact that 
$2\beta \cos(\pi f_a)$ plays the role of $\alpha$ to 
find 
\begin{equation} 
\Delta t= \frac{\pi t}{\hbar} 
\frac{\partial S_1}{\partial \alpha} (2\beta \sin{\pi f_a}) 
\,  \delta f_a \,. 
\end{equation} 
 
This allows us to write $s_2=\eta t \sqrt{E_J/E_c}$ 
where $\eta$ is of the order of unity. For the operating point 
we find $\eta \sim 3.5$. Therefore, changes in $H$ due to 
changes in $t_1$ go like $\sigma_x$. 
These tight-binding estimates for  
$\beta=0.8$ and $f_a=1/3$ give 
$s_1=0$ and $s_2=0.03$.


\begin{thebibliography}{99}
\bibitem{Lik-L} In fact, Likharev has pointed out that replacing the 
third junction by an inductor also will give a similar functional form 
for $U$. We choose the three junction system instead because the 
inductor values needed are not as accessible with our technology. We 
have also carried out quantum mechanical simulations with the inductor 
and do indeed find that it has similar properties to what is reported 
here.
\bibitem{H_footnote}Note that $H_t$  
could be have been found directly  
by not making the Legendre transformation 
in the Lagrangian. Then  the canonical momentum would not be 
proportional to the charges on the nodes. Although this canonical 
momentum and Hamiltonian are clearer mathematically,  
we  use instead  the more physical  canonical momentum  which  
proportional to the charge so as to make better contact 
with the literature, such as in Refs.~\cite{devoret-lecture}, 
\cite{Schoen88}, and \cite{Spiller92}.
\end{thebibliography}
\end{document}